\def\la{\hbox{{\lower -2.5pt\hbox{$<$}}\hskip -8pt\raise
-2.5pt\hbox{$\sim$}}}
\def\ga{\hbox{{\lower -2.5pt\hbox{$>$}}\hskip -8pt\raise
-2.5pt\hbox{$\sim$}}}
\def\ltsima{$\; \buildrel < \over \sim \;$}
\def\simlt{\lower.5ex\hbox{\ltsima}}
\def\gtsima{$\; \buildrel > \over \sim \;$}
\def\simgt{\lower.5ex\hbox{\gtsima}}

\documentstyle[epsfig]{elsart}

\begin{document}
\begin{frontmatter}
\title{On the Detectability of Gamma Rays from Clusters of Galaxies: Mergers
versus Secondary Infall}

\author[inaf2]{Stefano Gabici\thanksref{corr1}} 
\author[inaf1,infn]{Pasquale Blasi\thanksref{corr2}}
\address[inaf2]{Dipartimento di Astronomia, Universit\`a di Firenze,\\
Largo E. Fermi, 5 - I50125 Firenze, Italy}
\address[inaf1]{INAF/Osservatorio Astrofisico di Arcetri\\
Largo E. Fermi, 5 - I50125 Firenze, Italy}
\address[infn]{INFN, Sezione di Firenze, Italy}
\thanks[corr1]{E-mail: gabici@arcetri.astro.it}
\thanks[corr2]{E-mail: blasi@arcetri.astro.it}

\begin{abstract} 
Particle acceleration in clusters of galaxies is expected to take place
at both merger and accretion shocks. The electron component may be 
energized to energies of several TeV, and upscatter a small fraction of
the photons in the cosmic microwave background up to gamma ray energies. 
We address here the issue of the detectability of the gamma radiation 
generated either in merger events or during the accretion of matter onto 
cluster potential wells. The predictions are specialized to the cases of GLAST 
and AGILE, for which a few tens of clusters are expected to be detected, 
and to Cherenkov telescopes, for which however the perspectives do not 
appear to be optimistic. 
\end{abstract}
\end{frontmatter}

\section{Introduction}\label{sec:intro}

Observations of radio \cite{gigia} and hard X-ray \cite{fusco} emission 
from several clusters of
galaxies promoted these objects to the role of largest nonthermal sources
in the universe. On the theoretical side, two ideas contributed to reinforce
the belief that clusters of galaxies can be interesting as high energy 
radiation emitters: in \cite{bbp,volk} it was first understood that cosmic
rays accelerated within the cluster volume would be confined there for 
cosmological time scales, enhancing the possibility of inelastic 
proton-proton collisions and consequent gamma ray production through the 
decay of neutral pions. In \cite{wax1} it was recognized that large scale 
shocks associated with the formation of structures in the universe, may 
accelerate electrons to TeV energies, implying that high energy emission 
would occur due to the upscatter of the photons of the cosmic microwave 
background (CMB) to gamma ray energies, through inverse Compton scattering 
(ICS).

The possibility that single clusters could be sources of high energy
gamma rays fueled much interest also in the related possibility that 
clusters could in fact be also responsible for the observed isotropic 
diffuse gamma ray background (DGRB) \cite{egretdif}, thought to be of 
extragalactic origin (although the possibility that at least a fraction 
of this background might in fact be explained in terms of ICS from
electrons in the halo of our Galaxy \cite{dar}). 
In \cite{colablasi} it was found that the diffuse background due to 
inelastic interactions of cosmic rays trapped in the intracluster 
medium could not add up to more than $2\%$ of the observed flux above
$100$ MeV, confirming previous calculations illustrated in \cite{bbp}. The 
acceleration of electrons at the shocks related to the
process of structure formation was more recently considered in 
\cite{wax1}, where the observed DGRB was found to be saturated by the cluster 
contribution. A more realistic calculation \cite{gabici2} showed that 
ICS of electrons in clusters cannot contribute more than $\sim 10\%$ of 
the observed DGRB (of this, only $\sim 1\%$ is actually related to cluster
mergers, while most of the flux comes from electrons accelerated at accretion 
shocks). This result is in closer agreement with the results obtained in 
recent numerical calculations \cite{wax2,miniatifondo}. 

Although it seems now unlikely that clusters may play an important role
in explaining the bulk of the extragalactic DGRB, it seems plausible that
future gamma ray telescopes, especially in space, may finally be able to 
detect the gamma ray glow associated with the formation of 
single clusters of galaxies. In this paper we analyze this possibility 
more quantitatively, making use of a formalism developed in 
\cite{gabici1,gabici2} 
that allows us to evaluate the strength of the shocks that develop during 
mergers of two clusters, or during accretion of new matter onto an already 
existing cluster potential well. The important point made in \cite{gabici1} 
is that most merger related shocks are weak and unable to accelerate 
particles efficiently, a result later confirmed in \cite{dermer} and
in numerical simulations \cite{lastjones}.

Our calculations show that a gamma ray telescope like GLAST should 
be able to detect $\sim 50$ clusters of galaxies at energies above $100$
MeV. It is interesting that several clusters are expected to be detected
even with AGILE, while we confirm the undetectability of these sources
with EGRET. This is consistent with the recent results in \cite{olaf}
where new upper limits on the gamma ray emission from Abell clusters 
have been found, using old EGRET data.

The situation looks much less promising for ground based Cherenkov experiments,
which are optimized for point sources: nearby clusters, with larger fluences,
appear as relatively extended sources for these experiments, so that the 
corresponding potential for detection is lowered. Clusters at larger 
distances are point sources, but their fluence is well below the
detection threshold of Cherenkov experiments. Moreover, at energies in 
excess of a few hundred GeV the effect of absorption in the cosmological 
infrared background becomes important, and contributes to further suppress
the flux of gamma radiation from distant clusters. It is therefore unlikely 
that clusters of galaxies may be seen as shining at $100-1000$ GeV 
energies in the near future.

The paper is planned as follows: in \S \ref{sec:shock} we describe the 
acceleration of particles at shocks occurring during structure formation.
In \S \ref{sec:structure} we summarize our recipe for the calculation of
the strength of the shocks associated with the process of structure formation.
The consequences on the spectra of shock accelerated particles are also
summarized. In \S \ref{sec:space} we assess the detectability of clusters 
of galaxies with space-borne experiments, specializing our predictions to
the upcoming AGILE and GLAST satellites. In \S \ref{sec:cherenkov} the 
issue of detectability is discussed in connection with ground based
Cherenkov experiments. We conclude in \S \ref{sec:conclude}.

\section{Shock acceleration during structure formation}\label{sec:shock}

The hot gas in the intracluster medium is the proof that structure formation
proceeds through shock waves, that heat up the gas to the virial temperature
whose value is determined by the depth of the dark matter potential well.
The conversion of gravitational energy of the dark matter into thermal 
energy of the gas component occurs when the gas crosses shock waves that
are created during the process of structure formation. 

Diffusive acceleration can take place at these shocks, so that a small 
fraction of the particles in the accreting gas is extracted from the 
thermal distribution and energized to nonthermal energies. The acceleration
time, necessary to energize a particle to energy $E$ is given by \cite{lagage}:
\begin{equation}
\tau_{acc}(E) = \frac{3}{u_1-u_2} \left[\frac{D_1(E)}{u_1}+\frac{D_2(E)}{u_2}
\right],
\end{equation}
where $u_1$ and $D_1(E)$ ($u_2$ and $D_2(E)$) are the fluid speed 
and diffusion coefficient of the particles upstream (downstream). 
From the physical point of view, $\tau_{acc}$ is the sum of the residence 
times of particles upstream (index 1) and downstream (index 2). For well
behaved diffusion coefficients, larger magnetic field strengths correspond
to lower values of the diffusion coefficient, namely particles diffuse 
more slowly when the magnetic field is larger. This immediately suggests
that the diffusion coefficient upstream is larger than the same quantity
(at the same particle energy) downstream, so the residence time of the 
particles upstream is larger than the residence time in the downstream
region. The acceleration time can therefore be approximated as follows:
\begin{equation}
\tau_{acc}(E) \approx \frac{3 D_1(E) r}{u_1^2 (r-1)} = 
\frac{4 D_1(E)}{u_1^2} \left\{\frac{{\cal M}^2}{{\cal M}^2 - 1}\right\}, 
\label{eq:tauacc}
\end{equation}
where we introduced the compression factor $r=u_1/u_2$ and we used the 
relation between $r$ and the Mach number $\cal M$, $r=4{\cal M}^2/
({\cal M}^2+3)$. It is worth noticing that the acceleration time depends
in first approximation only on quantities that refer to the upstream fluid,
where the magnetic field has not been affected yet by the passage of the
shock, and preserves therefore its pre-shock structure. One may argue that
the diffusion coefficient in the downstream region is dominated by the 
scattering of particles on magnetic field fluctuation lengths comparable 
to the Larmor radius of the particles, namely $D_2(E)\approx (1/3) c E/e B_2$.
It is likely that these fluctuations are in fact created by the accelerated
particles themselves. In the upstream region however it is possible that 
the diffusion coefficient is determined by the magnetic field structure 
pre-existing the shock transit. In principle the momentum 
and magnetic field dependence of the two fields upstream and downstream 
may be different. 

For simplicity and convenience we assume here that the diffusion coefficient
on both sides is B\"{o}hm-like. Moreover, since the term in brackets in Eq.
\ref{eq:tauacc} varies between $1.8$ and $1$ when the Mach number changes
between $1.5$ and infinity, we neglect this term for the purpose of estimating
the acceleration time, therefore the acceleration time becomes:
\begin{equation}
\tau_{acc}\approx 0.3 B_\mu^{-1} E(GeV) \left(\frac{u_1}{10^8 \rm{cm/s}}
\right)^{-2}\rm years.
\end{equation}
We stress again here that the magnetic field $B_\mu$ and velocity $u_1$ 
refer to the unshocked medium. Here we focus our attention on electrons.
Their energy losses are dominated by ICS, with time scale:
\begin{equation}
\tau_{ICS}=\frac{E}{\frac{4}{3}\sigma_T c U_{CMB} 
\left(\frac{E}{m_e c^2}\right)^2} = 10^9 E(GeV)^{-1} \rm years, 
\end{equation}
where $U_{CMB}$ is the energy density in the CMB radiation. Equating the
acceleration and losses time scales we obtain an estimate of the maximum 
energy of accelerated electrons in case of B\"{o}hm diffusion:
\begin{equation}
E_{max}\approx 57 B_\mu^{1/2} \left(\frac{u_1}{10^8 \rm{cm/s}}\right)
~TeV.
\label{eq:Emax}
\end{equation}
For electron energies up to $\sim 400$ TeV the ICS off the CMB photons
occurs in the Thomson regime, therefore the maximum energy of the radiated 
photons is simply:
$$E_{\gamma,max} = \frac{4}{3} \left(\frac{E}{m_e c^2}\right)^2 \epsilon_{CMB}
\approx 7.5 TeV B_\mu \left(\frac{u_1}{10^8 \rm{cm/s}}\right)^2.
$$
As shown in \cite{gabici1,gabici2}, if the diffusion coefficient in the 
unshocked medium is not B\"{o}hm-like, then the maximum energy of the 
electrons may be appreciably lower than found in Eq. (\ref{eq:Emax}), and 
insufficient to upscatter the CMB photons to gamma ray energies. 
In the following we assume that the maximum energy is well defined by
Eq. (\ref{eq:Emax}). 

The value of the magnetic field to adopt in our calculations depends 
crucially on whether we are describing particle acceleration at merger
related shocks, or at accretion shocks. The former have relatively low
Mach numbers \cite{gabici1} and occur in the virialized regions of 
clusters of galaxies, where the magnetic field is expected to be in the
$\mu G$ range. Accretion shocks propagate in a cold unshocked medium,
where the magnetic field is expected to set at the cosmological value,
for which only upper limits are available. The field in the intracluster
medium could be the result of magnetic pollution from sources within 
the cluster \cite{kron}, being therefore unrelated to the magnetic field in 
the intergalactic medium outside clusters. Alternatively, the intracluster
magnetic field could result from the compression of the external field,
if the latter is of cosmological origin. The energy of the photons radiated
by the accelerated electrons through ICS is above $E_\gamma$ whenever the 
magnetic field $B_\mu$ is larger than $10^{-9} (u_1/1000 km/s)^{-2}(E_\gamma/
10 GeV)$ Gauss.
This bound is certainly satisfied by the intracluster magnetic field, 
not necessarily by the poorly known field in the intergalactic medium
outside clusters, for which upper limits of $10^{-9}-10^{-11}$ have been
found, depending on the field structure \cite{burles}. On the other
hand this constraint has to be taken with caution for the case of accretion
shocks, because in a more realistic situation, as suggested by numerical 
simulations, high Mach number shocks are located close to filaments, where
a substantial compression of cosmological fields might have occurred.
We assume in the following that the electrons accelerated at both merger
and accretion shocks are energetic enough to radiate gamma ray photons
up to the $\sim TeV$ region.

In the next section we will show that there is a 
wide spread in the Mach numbers from about unity to $\sim 1000$ 
\cite{miniatishocks}, with a peak around $1.5$ \cite{gabici1} for the 
weak merger related shocks.
There is a correspondingly large spread in the power law index of the 
accelerated particles, while very little spread exists in the maximum energy 
that can be achieved at these shocks.

\section{Structure formation and strength of the merger and accretion shocks}
\label{sec:structure}

It is commonly believed that clusters of galaxies form hierarchically, with 
smaller dark matter halos merging together to form more massive halos. The
shock surfaces that form during these mergers are responsible for the 
conversion of gravitational energy into thermal energy of the intracluster
medium. Parallel to the merger processes, a secondary infall \cite{bert} of 
gas occurs at all times onto the potential well which is being formed. The 
information about the virialization of the inner region of this accretion 
flow is propagated outwards through a so-called {\it accretion shock}. While 
merger related shocks form and propagate in the virialized gas, accretion 
shocks propagate in the cold ($T\sim 10^4-10^6 K$) intergalactic medium, 
and therefore have, almost by definition, very large Mach numbers. In first
approximation the position of the accretion shock is roughly coincident with
the virial radius of the cluster.

In \cite{gabici2} we introduced a clear distinction between merger related
shocks and accretion shocks, as seen from the perspective of nonthermal 
phenomena rather than from the point of view of structure formation.
This classification however turns out to be useful for structure formation
studies as well \cite{lastjones}.

A useful analytical description of this hierarchical clustering was proposed
by Press and Schechter (PS) \cite{press} and subsequently developed, among 
others, by Lacey and Cole \cite{laceyCole} in the form of the so-called 
extended Press-Schechter formalism. 
Within this approach, the comoving number density of clusters with mass 
$M$ at cosmic time $t$ is given by:
\begin{equation}
\frac{dn(M,t)}{dM}=\sqrt{\frac{2}{\pi}}\,\frac{\varrho}{M^2}\,
\frac{\delta_c(t)}
{\sigma(M)}\,\left|{\frac{d\ln \sigma(M)}{d\ln M}}\right| exp\left[-\frac
{\delta_c^2(t)}{2\sigma^2(M)}\right],
\label{eq:ps}
\end{equation}
while the rate at which clusters of mass $M$ merge at a given time to form 
clusters with final mass $M^{\prime}$ is:
$${\cal R}(M,M^{\prime},t)dM^{\prime}=$$
\begin{eqnarray}
\sqrt{\frac{2}{\pi}}\,\left|
\frac{d\delta_c(t)}{dt}\right|\,\frac{1}{\sigma^2(M^{\prime})}\,
\left|\frac{d\sigma(M^{\prime})}{dM^{\prime}}\right|\,
\left(1-\frac{\sigma^2(M^{\prime})}{\sigma^2(M)}\right)^{-3/2} \nonumber \\
\rm{exp}\left[-\frac{\delta_c^2(t)}{2}\left(\frac{1}{\sigma^2(M^{\prime})}-
\frac{1}{\sigma^2(M)}\right)\right]dM^{\prime}.
\end{eqnarray}
Here $\varrho$ is the present mean density of the universe, $\delta_c(t)$ 
is the critical density contrast linearly extrapolated to the present time 
for a region that collapses at time $t$, and $\sigma(M)$ is the current rms 
density fluctuation smoothed over the mass scale $M$ (see \cite{nippo1,nippo2} for more details). 

As discussed above, shock waves form naturally in the
baryonic component of clusters of galaxies involved in a merger, due to 
the supersonic motion of the two clusters. In \cite{gabici1} we developed 
an analytical recipe that allows us to estimate the Mach number of merger 
related shocks, once the masses of the two merging clusters and the redshift 
where the merger event occurs are given. The virial radii of the two 
clusters are written in the form
$$
R_i = \left(\frac{GM_i}{100\Omega_mH_0^2(1+z_{f,i})^3}\right)^{1/3},
$$
where $\Omega_m$ is the fraction of matter in the universe, compared to
the critical density, $H_0$ is the Hubble constant and $z_{f,i}$ is the 
formation redshift of the $i$-th cluster.
The relative velocity $v$ of the two merging clusters, with mass $M_1$ and 
$M_2$ respectively can be evaluated from energy conservation:
\begin{equation}
-\frac{GM_1M_2}{R_1+R_2}+\frac{1}{2}M_rv_r^2 = -\frac{GM_1M_2}{2R_{ta}},
\end{equation}
where $M_r$ is the reduced mass, and $R_{ta}$ is the turnaround radius 
where the clusters are supposed to be at rest with respect to each other.
In an Einstein-De Sitter cosmology the turnaround radius is equal to twice 
the virial radius of the system and this result holds in approximate way also 
for a flat, lambda cosmology \cite{lambda}. The virial radius depends on the 
cluster formation redshift, that we estimate following \cite{laceyCole}.

The sound speed in a cluster of mass $M_i$, needed in order to evaluate 
the Mach numbers, can be written as follows:
$$
c_{s,i}^2 = \gamma (\gamma-1) \frac{GM_i}{2R_i}~,
$$
where $\gamma = 5/3$ is the adiabatic index of the gas.

\begin{figure}
\begin{center}
\includegraphics[width=0.7\textwidth]{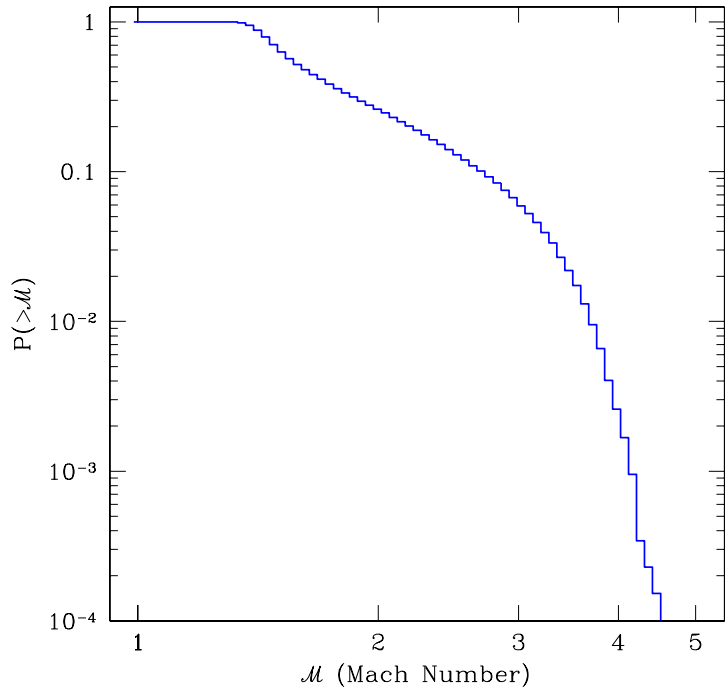}
\caption{Cumulative distribution of Mach numbers in mergers of clusters of
galaxies.}
\label{fig:machs}
\end{center}
\end{figure}

Using this recipe, we can determine the probability of having a merger shock
with a given Mach number, simply by simulating the merger tree of a cluster
with given mass today. The cumulative probability distribution of shock Mach 
numbers, as obtained in \cite{gabici1}, is plotted in Fig. \ref{fig:machs}.
It is clear from the figure that most merger related shocks have Mach number
around $\sim 1.5$. As a consequence, the spectra of particles accelerated 
in most merger related shocks are predicted to be extremely steep, and 
therefore of marginal importance from the point of view of explaining the 
observed nonthermal phenomena in clusters of galaxies. On the other hand, the
strong shocks expected in the case of mergers between clusters with very 
different masses may generate periods of large gamma ray emission 
lasting approximately for the same duration of the merger itself. 
This is expected to happen in $\sim 5\%$ of the cases, as derived from 
Fig. \ref{fig:machs} \cite{gabici1}.

During the mergers, a population of newly accelerated protons is also 
injected in the intracluster medium.
These protons can contribute high energy gamma rays through the decay 
of neutral pions and high energy electrons through the decay of charged 
pions \cite{bbp,ensslin,colablasi,blasicola,blasi99,blasi01}.
As pointed out in \cite{dermer}, an hadronic feature might appear on top
of the gamma ray emission generated by ICS of electrons, even 
if protons are accelerated with the same efficiency as electrons.

However it is important to keep in mind that the proton component
cannot escape the cluster volume, therefore the energy density in accelerated
protons in a cluster depends wildly upon the merger history of the cluster,
and on the microphysics of the shock acceleration process, as will be
discussed in a forthcoming paper. This is mainly a consequence of the 
re-energization of the protons stored in the intracluster medium each 
time that a new merger event takes place. This problem deserves a 
detailed discussion {\it per se}, and we decided to dedicate a separate
paper to this issue.

It may be useful to summarize the content of this section as follows:
structure formation occurs through hierarchical clustering, consisting
of mergers of smaller clusters, which provide the largest contribution 
to the mass enhancement, and of accretion of matter onto the evolving 
potential well. 
Both processes are characterized by the formation of shocks with very 
different properties: merger shocks are mostly weak and occur in the
central parts of clusters, while accretion shocks form in the outskirts
and have large Mach numbers. The results about merger shocks, obtained
in \cite{gabici1} were initially discrepant to the results of numerical
simulations \cite{miniatishocks}, which found a bimodal distribution of Mach
numbers with two peaks at Mach numbers $\sim 5$ and $1000$. More
recent numerical simulations \cite{lastjones} found that the peak
at lower Mach number is in fact at $1.5$ as previously obtained 
with our semi-analytical calculations \cite{gabici1}. The peak at larger
Mach numbers in \cite{lastjones} is simply due to accretion shocks. These
shocks were not included in our initial paper \cite{gabici1} and 
were introduced later in a separate pubblication \cite{gabici2}.

\section{Detectability of gamma rays from clusters of galaxies with 
space-borne gamma ray telescopes}\label{sec:space}

As discussed in many previous papers, gamma rays can be generated in 
clusters of galaxies mainly through the following physical processes 
(see e.g. \cite{blasi01}):
\begin{itemize}
\item[1.] ICS of electrons accelerated at merger and accretion shocks
\item[2.] ICS of secondary electrons from the decay of charged pions produced
in inelastic $pp$ collisions 
\item[3.] Decay of neutral pions produced in inelastic $pp$ collisions
\item[4.] Bremsstrahlung emission of relativistic primary and secondary 
electrons.
\end{itemize}
Due to cosmic ray confinement, protons are stored in the intracluster
medium for cosmological times, and in principle can produce secondary 
products continuously \cite{bbp}, namely with no time correlation with 
phenomena like mergers of two clusters. While this process is of great 
importance and can be responsible for appreciable gamma ray emission,
it is also quite uncertain because the amount of protons diffusively
stored in the intracluster medium is determined by the history of the
cluster. In the following we discuss only three of the elements that 
contribute to this uncertainty:

{\it Mergers as accelerators}

Different clusters have different merger histories, namely different 
merger trees. As stressed above, most mergers produce weak shocks, 
unable to efficiently energize cosmic ray particles. Depending on the
cluster, there may be several strong shocks formed when mergers between 
clusters with very different masses take place. The energy density in protons
trapped within the cluster that will eventually result from several merger
events occurring during its history depends on the number and strength 
of the shocks that accelerated those protons. 

{\it Mergers as re-accelerators}

While most electrons accelerated during a merger lose most of their energy 
before the next merger event, protons are simply stored in the intracluster 
medium and each new merger related shock re-energizes the particles trapped
until that time, besides accelerating new particles. Again, the merger history 
affects this component in a crucial way. Moreover, the spectrum of the 
re-energized particles depends on the spectrum of the protons accelerated
and re-energized by previous mergers. The spectrum of the re-accelerated
particles also depends upon the minimum energy that the protons possess.
These effects may change the spectrum of the accelerated particles trapped in 
the intracluster medium even by orders of magnitude, in particular at the
highest energies, where the spectrum depends crucially on the amount of
energy that crossed the few strong shocks developed during the merger
history of the cluster. A corresponding uncertainty affects
the secondary products of cosmic ray interactions in the intracluster
medium, namely gamma rays and electrons (positrons).

{\it Additional sources}

The storage properties that clusters of galaxies exhibit for the proton
component make them very sensitive to all nonthermal events occurring
within the cluster. In particular, acceleration processes in radio galaxies
in cluster cores pollute the intracluster medium with accelerated protons
which have no connection with shocks related to structure formation. 

All these uncertainties can be seen as pieces of physics of the cosmic
ray acceleration in clusters of galaxies that require further investigation,
and will be discussed at length in a forthcoming publication. 
In the present paper on the other hand we adopt a conservative attitude
and restrict our attention to the gamma ray emission that is solely the
result of ICS of relativistic electrons, accelerated at merger and accretion
shocks, against photons of the CMB radiation. Any contribution to the
gamma ray emission from proton interactions can only increase the gamma
ray fluxes derived below.

We start our discussion from accretion shocks, whose role is simpler to 
investigate since their Mach number is extremely large and the spectrum
of the particles they accelerate is almost exactly a power law $E^{-2}$.
The balance between energy losses, dominated by ICS, and continuous injection 
of newly accelerated electrons, drives the injection spectrum towards an 
equilibrium spectrum that is one power steeper than the injection. 

The energy per unit time converted at the shock into nonthermal electrons
can be written as
\begin{equation}
L_e^{acc} = \eta~ \frac{1}{2}~ \varrho_b~ (1+z)^3~ v_{ff}^3~ 4 \pi~ R_v^2~  
\propto ~M^{5/3}
\end{equation}
where $\eta_e$ is the shock acceleration efficiency for electrons, fixed
here at $5\%$, $v_{ff}=(2GM/R_v)^{1/2}$ is the free fall velocity of the 
gas at the virial radius $R_v$ and $\varrho_b=\Omega_b\rho_{cr}$ is the 
present mean baryonic density of the universe, if $\Omega_b=0.02h^2$ is the 
baryon fraction and $\rho_{cr}$ is the critical density.

It is important to keep in mind that the accretion of gas onto the cluster
only contributes a small fraction ($\sim 10\%$) of the total final mass, 
which is instead mainly built up through mergers with other massive 
clusters. 

The differential gamma ray luminosity, expressed in photons per unit energy
per unit time, is easily calculated as ICS of the CMB photons, and is 
proportional to $L_e^{acc}$ given above. It follows that, for a given redshift, the 
gamma ray luminosity is a growing function of the cluster mass, so that the 
number of accreting objects observable by a gamma ray telescope with 
sensitivity $F_{lim}$ can be written as follows:
\begin{equation}
N_{acc}(F_{lim}) = \int^{\infty}_0 dz \frac{dV}{dz} 
\int^{\infty}_{M(F_{lim},z)} dM n(M,z)~,
\end{equation}
where $dV$ is the comoving volume in the redshift region between $z$ and 
$z+dz$, $n(M,z)$ is given 
by Eq. \ref{eq:ps} and $M(F_{lim},z)$ is the mass of a cluster accreting 
at redshift $z$ whose flux is $F_{lim}$.

The case of merger shocks is slightly more complicated, due to the 
fact that the shock strength is a function of the masses of the clusters
involved in the merger, and more specifically of the ratio of the masses. 
As discussed above, merger shocks are often weak, following the distribution 
given in Fig. \ref{fig:machs}. Flat electron spectra, necessary to generate 
an appreciable gamma ray emission above $100$ MeV, are obtained at strong 
shocks that form only during mergers between clusters with very different 
masses. 
In the approach presented in \cite{gabici2}, at each merger events two
shocks are generated, each propagating in one of the merging clusters.
The Mach numbers of both shocks can be calculated from the relative 
velocity of the clusters and if the sound speed in each cluster is known. The
slopes of the spectra of the accelerated electrons easily follows as
$\alpha = 2({\cal M}^2+1)/({\cal M}^2-1)$, where ${\cal M}$ is the relevant
Mach number.

The rate at which energy is channelled into nonthermal electrons in 
merger related shocks can be evaluated as 
$$
L_e^{mer} = \eta \frac{1}{2}\varrho_{b,i} v^3 S_i 
$$
where $\varrho_{b,i}$ is the baryonic density of the i-th cluster, $S_i$ is 
the surface of the shock that propagates in the i-th cluster and $v$ is 
the merger relative velocity.
The gamma ray luminosity due to ICS of these accelerated electrons is 
calculated as usual.

The number of observable merging clusters is:
$$
N_{mer}(F_{lim}) = \int^{\infty}_0 dz \frac{dV}{dz} \int^{\infty}_{M_{min}} 
dM_1 n(M_1,z)\times
$$
\begin{equation}
\int^{M_1}_{M_{min}} dM_2 {\cal R}(M_1,M_1+M_2,z)\Delta t_{mer} 
\vartheta[F_{lim}-F_{\gamma}(M_1,M_2,z)],
\end{equation}
where $F_{lim}$ is the telescope sensitivity, $F_{\gamma}(M_1,M_2,z)$ is the 
gamma ray flux we receive when a cluster with mass $M_1$ merges with a cluster
with mass $M_2$ at a redshift $z$, $\Delta t_{mer}=R_2/v$ is the duration 
of the merger event, $M_{min} = 10^{13} M_{\odot}$ is the typical mass for 
galaxy groups and $\vartheta$ is the Heaviside step function. 
\begin{figure}
\begin{center}
\includegraphics[width=0.7\textwidth]{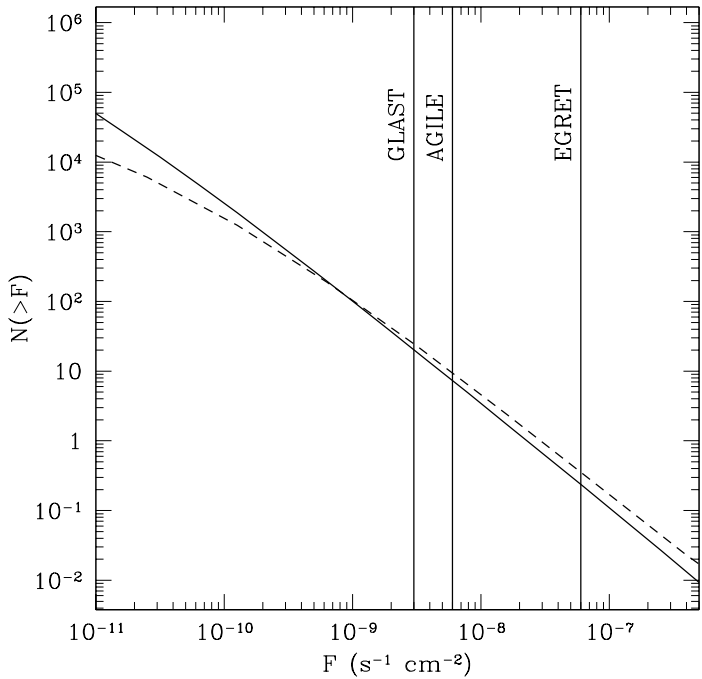}
\caption{Number of accreting (solid line) and merging (dashed line) clusters 
with gamma ray flux greater than $F$. The vertical lines represent the GLAST, 
AGILE and EGRET sensitivity for point sources.}
\label{fig:counts}
\end{center}
\end{figure}
The aim of this section is to determine the $\log N-\log S$ distribution 
of clusters of galaxies as gamma ray sources, for the cases of 
mergers and accretion. The results of our calculations are shown in 
Fig. \ref{fig:counts} in the form of number of clusters with gamma ray
emission at energies in excess of 100 MeV larger than some value $F$ on
the x-axis: the solid and dashed lines refer to accretion and mergers 
respectively. The vertical lines correspond to the GLAST, AGILE and EGRET 
sensitivities for point sources, as indicated in the plot. 

The fact that the two curves in Fig. \ref{fig:counts} overlap almost
exactly is the result of a combination of many factors: each 
merger is on average much more luminous that accretion, because the 
energy involved is larger and the time during which it is released
is shorter ($\sim 10^9$ years for a merger versus the cluster lifetime
for accretion). On the other hand, most of the times the merger event
does not result into gamma ray emission, because the shocks involved
are too weak. The slight deviation from a power law at low fluxes 
in Fig. \ref{fig:counts} is the consequence of the 
cosmological effects in a non-euclidean geometry. The change of slope 
appears at larger fluxes for the mergers (dashed line) than for accretion.
This is a result of the larger luminosity per merger that allows one to 
detect a bright merger from larger distances but also the consequence of the
strong evolution in the merger luminosity. This latter effect is easy to 
explain: for a minimum cluster mass of $\sim 10^{13}M_\odot$, the flat
spectra are obtained only when a merger occurs with a cluster of mass
$\sim 10^{15}M_\odot$; these clusters are already on the tail of the PS 
distribution at the present cosmic time, while they had not been formed
yet at much earlier times. Despite the larger luminosity of the mergers 
(on average), it is interesting to notice that the gamma ray emission for
the merger case is limited to the merger period or shortly after, while 
the gamma ray emission associated to accretion is continuous, although 
less intense.  

Our calculations show that $\sim 50$ clusters should be detected in 
gamma rays above 100 MeV by GLAST, equally distributed between merging
and accreting clusters. AGILE on the other hand should be 
able to detect $\sim 10-20$ objects. Moreover, we predict that no cluster 
should have been detected by EGRET, despite the recent claims of 
temptative association of some unidentified EGRET sources with the 
location of Abell clusters \cite{cola,gg}. 

The results summarized in Fig. \ref{fig:counts} are compared with 
the sensitivity of EGRET, GLAST and AGILE as obtained for point sources. 
This seems justified at least for gamma rays with energy around 100 MeV: 
the gamma ray emission region in clusters is approximately a few degrees 
wide and the point spread function of GLAST at 100 MeV is $\sim 4^o$,
while it reduces to $0.1^o$ at 10 GeV \cite{web}. Similar arguments hold 
even for the AGILE angular resolution \cite{web2}.

Since our results greatly differ from some previous results, some 
comments and comparisons are required. In \cite{totkita,waxlo} 
a Press-Schechter based method similar to the one presented in 
\cite{gabici2} was developed 
in order to make predictions about the number of clusters detectable in
gamma rays with energy above $100$ MeV with future gamma ray telescopes. 
The authors claimed that at least a few tens of clusters should have been 
visible to EGRET. GLAST, on the other hand, was predicted to be able to
detect more than a few thousands of such objects.
This prediction seemed to be supported by some preliminary observational 
evidences for a temptative association between unidentified high latitude 
EGRET sources and Abell clusters (or cluster pairs) \cite{cola,gg}. The 
statistical significance and physical plausibility of such an association 
was strongly questioned in \cite{olaf}, in which the important conclusion 
was reached that {\it we still have to await the first observational evidence 
for the high-energy gamma-ray emission of galaxy clusters}. 
The lack of association between unidentified EGRET sources and clusters
of galaxies was also found in \cite{scharf}. These findings seem to be
perfectly in line with the predictions described in the present paper.

In \cite{totkita}, all shocks were assumed to be strong. As a consequence,
the spectrum of the accelerated particles was always taken to be $\propto
E^{-2}$. As discussed above, and as described in greater details in 
\cite{gabici1,gabici2}, for merger related shocks this assumption leads
to incorrect results. The gamma ray emission in \cite{totkita} is 
overestimated by orders of magnitude (despite the fact that the fraction of
the energy crossing the shock which is converted to nonthermal electrons
is assumed to be the same as here, namely $5\%$), and its spectrum does not 
reflect the real strength of the shocks developed during mergers of clusters
of galaxies. This concept can be rephrased in another way: at fixed total
energy flux crossing the merger shocks, and at fixed efficiency of electron
acceleration, the nonthermal energy is distributed among accelerated particles
in a different way if the spectrum of nonthermal electrons is taken to be
$E^{-2}$ rather than as the one derived from the correct evaluation of the 
shocks strength. The clusters that we predict to be 
detectable with AGILE and GLAST in gamma rays above 100 MeV all have flat
spectra, which means that the gamma ray emission selects the few mergers
that happen to have a flat spectum of accelerated electrons (note that 
for Mach numbers in the range $1.5-4$, the slope of the electron spectrum
spans the range $\alpha \sim 2-5$), in addition to the less luminous nearby 
accreting clusters that always generate flat spectra.

Similar arguments apply to the calculations in Ref. \cite{waxlo}: 
the formation of a cluster was exemplified there as a spherical collapse,
in which $\sim 5\%$ of the kinetic energy crossing the accretion shock
was converted into nonthermal electrons with spectrum $E^{-2}$. From
the energetic point of view, this approach is roughly equivalent to 
that of \cite{totkita}. However, as discussed above, only a relatively small
fraction of the mass in a cluster is expected to be accumulated through 
accretion, being mergers with other clusters the main physical process
for mass build-up. As stressed above, the spectrum of the particles that
the energy is channelled into is the key point in understanding 
whether one can expect appreciable gamma ray fluxes from clusters of
galaxies. 

\section{Detectability of clusters with ground based Cherenkov telescopes}
\label{sec:cherenkov}

In \S \ref{sec:shock} we showed that if electrons are accelerated
diffusively across merger and accretion shocks following B\"{o}hm diffusion, 
their maximum energy can be as high as several tens of TeV, and the CMB 
photons can be upscattered through ICS to gamma ray energies up to several TeV 
if the magnetic field in the vicinity of the shock is in the range 
$0.1-1~\mu G$. 

In this section we address the issue of detectability of the gamma
rays with energy above $100$ GeV from clusters of galaxies with Cherenkov 
ground based detectors. The calculations of the gamma ray emission for 
merging and accreting clusters have already been described in the previous 
section. The only new ingredient that needs to be included here 
is the absorption of high energy gamma rays due to pair production
in the universal photon background, in particular the infrared background 
(IRB). Pair production starts to be important when the energy in the center of
mass of a photon-photon scattering equals twice the electron mass. 
The net result of the propagation of high energy gamma rays with energy 
in excess of a few hundred GeV is an absorption cutoff for sources more
distant that the pathlength for pair production. The corresponding optical 
depth at energy $E$ for a source at distance $D$, $\tau(E,D)$, has been 
calculated in \cite{stecker} where the authors adopt an empirical model 
for the IRB, poorly measured in the energy range of interest. The gamma
ray flux calculated from one cluster in the previous section can be 
translated into a flux at the detector simply multiplying it by $\exp\left[
-\tau(E,D)\right]$. 

\begin{figure}
\begin{center}
\includegraphics[width=0.7\textwidth]{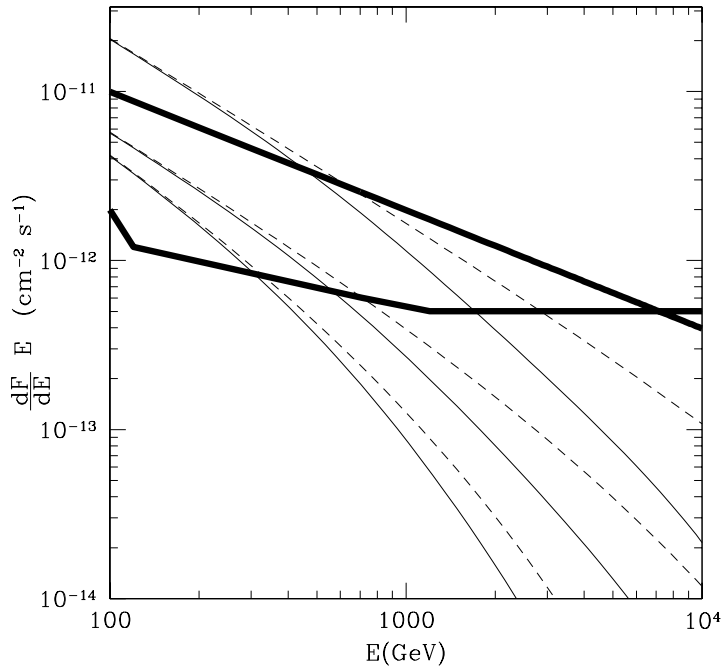}
\caption{Gamma ray emission in the 100 GeV - 10 TeV region. The thick solid
lines represent the sensitivities of a IACT for point sources (lower curve) 
and extended sources (upper curve). The predicted gamma ray fluxes from a 
Coma-like cluster at a distance of 100 Mpc with and without absorption of 
the infrared background are plotted as dashed and solid lines respectively.}
\label{fig:cherenkov}
\end{center}
\end{figure}

In Fig. \ref{fig:cherenkov} we plot the results of our calculations for the
high energy gamma ray spectrum generated from a Coma-like cluster of galaxies 
at 100 Mpc distance for the case of merger and accretion. The effect of the
gamma ray absorption in the IRB is illustrated by the difference between 
the solid lines (with absorption) and dashed lines (without absorption).

From top to bottom, the lines refer to three different cases: 1) a merger
between two clusters with masses $10^{15}M_\odot$ and $10^{13}M_\odot$; 2)
an accreting cluster with mass $10^{15}M_\odot$ with a magnetic field
at the shock in the upstream region $0.1\mu G$; 3) an accreting cluster with 
mass $10^{15}M_\odot$ with a magnetic field at the shock in the upstream 
region $0.01\mu G$. 

The thick solid lines represent the sensitivities for a generic Cherenkov 
telescope as calculated in \cite{felix}. These results are obtained 
considering an array of imaging atmospheric Cherenkov telescopes (IACT) 
consisting of $n$ cells, each consisting of a $100\times 100~\rm m^2$ 
quadrangle with four '100 GeV' class IACTs in its corners. The two thick 
curves in Fig. \ref{fig:cherenkov} represent the minimum detectable fluxes 
for point sources (lower curve) and an extended $1^{\circ}$ wide source 
(upper curve) for an exposure time of 1000 hours. The exposure time here 
is defined as the product between the 
observation time and the number of cells that form the array. For instance, 
an exposure of 1000 hours can be achieved with a 100 hours observation 
performed by an array consisting of 10 cells.
If the angular resolution of the Cherenkov experiment is taken to be
$0.1$ degrees, most clusters of galaxies, both merging and accreting,
are diffuse sources for these experiments. The angular size of 
a cluster at distance $D$ and with an emitting region of size $R=1$ Mpc
is $\alpha\approx 1^o~(D/100~\rm Mpc)$. For accretion, the shock surface
at which acceleration occurs is located approximately at the virial
radius, that for a Coma-like cluster is $\sim 3$ Mpc, therefore the
emission is even more extended than for a merger, and with angular
size that approaches or even exceeds the aperture of the Cherenkov
telescopes (a few degrees). 

For the optimistic conditions that Fig. \ref{fig:cherenkov} refers
to, the predicted flux of gamma rays between 
100 GeV and a few hundred GeV for a merger between two clusters with
mass ratio $10^{-2}$ is only slightly above the sensitivity of a 
Cherenkov telescope for a 1000 hours exposure. The predicted fluxes
from accretion are unobservable. The sensitivities of the telescopes
might be improved to some extent, compared with the curves in Fig.
\ref{fig:cherenkov} by simulating off-center showers and different
triggering modes of the telescopes in a cell. It appears however that 
the perspectives for detection of TeV gamma radiation from ICS of 
ultrarelativistic electrons in clusters of galaxies are not very 
promising. 

\section{Discussion and Conclusions}
\label{sec:conclude}

The detection of gamma ray emission from clusters of galaxies represents
one of the major goals of future gamma ray telescopes. In this paper we
addressed the issue of detectability of the gamma ray signal as expected
from mergers of clusters of galaxies and from the so-called secondary 
infall of gas onto a cluster potential well. We stressed that the two 
phenomena are intrinsically different both for the energy involved and for
the spectra of the particles accelerated at the related shock surfaces. 

Gamma rays can be produced mainly due to two physical processes, namely
generation and decay of neutral pions in inelastic $p-p$ collisions, and
inverse Compton scattering of relativistic electrons. In this paper we 
concentrated our attention on the latter process, while the full account
of the gamma ray emission from hadronic processes will be discussed in 
detail in a forthcoming paper. The reason for separating the two calculations
is the peculiar behaviour of protons (or more generally nuclei) that are
stored in the intracluster medium for cosmological times, causing the 
radiation produced at any fixed time to be the result of the full history
of the cluster. The situation is
made more difficult by the fact that shocks related to structure formation
re-energize the protons stored in the intracluster medium, besides
accelerating new cosmic rays from the thermal plasma. All these processes
depend on the specific history of the cluster, determined by the merger
tree that the custer experiences. Each shock associated to different 
mergers has different strength, therefore even the spectrum of the 
accelerated and re-accelerated particles depends on the merger history.
As a result, the energy density in cosmic rays in a cluster may change
wildly from cluster to cluster.

The situation is simpler for electrons accelerated during structure 
formation: the electrons responsible for the gamma ray emission through
inverse Compton upscattering  
of the CMB photons have energies in the range 200 GeV-50 TeV, whose
loss time scale is between thirty thousand years and a few million years,
much shorter than the time between two merger events. Most electrons
are therefore simply accelerated during a merger and lost to lower 
energies afterwards. 

Clusters of galaxies grow mainly through mergers, but secondary infall
of gas onto a cluster also contributes to the mass increase, although 
in a smoother way, without the billion-years-long bursts that are typical 
of mergers. 

The main differences between the gamma rays generated by particles accelerated
at mergers and accretion shocks are the energy available and the spectra
of accelerated particles.
As found in \cite{gabici1} and summarized here, merger related shocks are
mostly weak, their Mach number distribution having a peak at $M\sim 1.5$.
These shocks generate very steep electron spectra, irrelevant from the point
of view of high energy phenomena related to electrons. The only mergers able 
to produce flat spectra are the so-called minor mergers, consisting of the 
encounter of two clusters with mass ratio $\sim 0.01$. The gamma ray emission 
in these events is restricted to the duration of the merger and likely 
limited to the shock region.

The accretion shocks are located in the outskirts of clusters, at a 
position comparable with the virial distance from the cluster center. 
The energy flow per unit time across accretion shocks is smaller than
that related to mergers, but it occurs continuously in time. Since these shocks
carry the information about the virialization of the inner regions towards
the cold outer parts of the cluster, by definition their Mach number can
be very high, $M \sim 10-1000$. As a consequence, the spectrum of the 
particles accelerated at these shocks all have spectrum $E^{-2}$ (note
that the equilibrium spectrum, due to the balance between injection and
losses is one power steeper than the injection spectrum).

We calculated the $\log N-\log S$ distribution of merging and accreting
clusters of galaxies and compared the expected fluxes with the sensitivities
of EGRET, AGILE and GLAST. We found that no cluster was expected to be
detected by EGRET, confirming a recent analysis of the old EGRET data
from the direction of several Abell clusters, for which only upper limits
on the gamma ray emission could be inferred \cite{olaf}. Our calculation
strongly disagrees with previous results found in \cite{totkita,waxlo}, 
for the reasons explained in \S \ref{sec:space}.
 
The perspectives for detection appear to be more promising for future 
space-borne gamma ray 
telescopes, such as AGILE and GLAST. Our calculations show that $\sim 10-20$
clusters should be detected by AGILE and $\sim 50$ clusters should appear 
in GLAST data, equally shared between merging and accreting clusters if an
efficiency of $\sim 5\%$ can be achieved in accelerating electrons at the
shocks. 

In principle, the gamma ray emission due to ICS of ultrarelativistic electrons
can extend to a few tens of TeV, so that clusters might be seen in future 
ground based Cherenkov experiments such as VERITAS and HESS. Unfortunately,
only nearby ($\sim 100~\rm Mpc$) merging clusters of galaxies are expected
to be detectable, while the detection of accreting clusters appears more
problematic. 

\section*{Acknowledgments} We thank Gianfranco Brunetti for many 
useful discussions and ongoing collaboration. We are also grateful to
Felix Aharonian for constructive correspondence about Cherenkov experiments
and to Tom Jones and Dongsu Ryu for updating us on their work.
This work was partially supported through grant COFIN 2002 at the Arcetri 
Astrophysical Observatory.

\end{document}